\documentclass[manuscript,screen]{acmart}
\AtBeginDocument{%
  \providecommand\BibTeX{{%
    \normalfont B\kern-0.5em{\scshape i\kern-0.25em b}\kern-0.8em\TeX}}}

\usepackage{xspace}
\usepackage{mathrsfs}
\usepackage{algorithmic}
\usepackage{graphicx}
\usepackage{textcomp}
\usepackage{xcolor}
\usepackage{paralist}
\usepackage{hyperref}
\usepackage{lipsum}
\usepackage{url}

\newcommand{\npm}{\emph{npm}}

\newcommand{\Product}{\mathscr{P}}
\newcommand{\Revision}{\mathscr{R}}
\newcommand{\product}{\mathrm{product}}
\newcommand{\dep}{\mathrm{dep}}
\newcommand{\resolved}{\mathrm{resolved}}
\newcommand{\vint}{V^{\mathrm{int}}}
\newcommand{\vext}{V^{\mathrm{ext}}}

\usepackage{tikz}

\newlength{\RoundedBoxWidth}
\newsavebox{\GrayRoundedBox}
\newenvironment{GrayBox}[1][\dimexpr\columnwidth-4.5ex]%
   {\setlength{\RoundedBoxWidth}{\dimexpr#1}
    \begin{lrbox}{\GrayRoundedBox}
       \begin{minipage}{\RoundedBoxWidth}}%
   {   \end{minipage}
    \end{lrbox}
    \begin{center}
    \begin{tikzpicture}%
       \draw node[draw=black,fill=black!10,rounded corners,%
             inner sep=2ex,text width=\RoundedBoxWidth]%
             {\usebox{\GrayRoundedBox}};
    \end{tikzpicture}
    \end{center}}

\begin{document}

\title{Fine-Grained Network Analysis for Modern Software Ecosystems}
\titlenote{This is an extended version of the paper \emph{How Network Analysis Can Improve the Reliability of Modern Software Ecosystems} presented by the
first author in 2019 at the IEEE International Conference on Cognitive Machine Intelligence (CogMI). Sections 3, 5, 6 are new and the remaining parts have been largely reworked.}
\author{Paolo Boldi}
\email{paolo.boldi@unimi.it}
\affiliation{
	\institution{Dipartimento di Informatica, Universit\`a degli Studi di Milano}
	\city{Milan}
	\country{Italy}
}
\author{Georgios Gousios}
\email{g.gousios@tudelft.nl}
\affiliation{
  \institution{Department of Software Technology, Delft University of Technology}
  \city{Delft}
  \country{The Netherlands}
}
\renewcommand{\shortauthors}{Paolo Boldi and Georgios Gousios}

\begin{abstract}
Modern software development is increasingly dependent on components, libraries and frameworks coming from third-party vendors or open-source suppliers
and made available through a number of platforms (or \emph{forges}). This way of writing software puts an emphasis on reuse and on composition, commoditizing
the services which modern applications require. On the other hand, bugs and vulnerabilities in a single library living in one such ecosystem can affect, directly or
by transitivity, a huge number of other libraries and applications. Currently, only product-level information on library dependencies is used to contain this kind of danger,
but this knowledge often reveals itself too imprecise to lead to effective (and possibly automated) handling policies. We will discuss how fine-grained function-level
dependencies can greatly improve reliability and reduce the impact of vulnerabilities on the whole software ecosystem.
\end{abstract}



\maketitle

\section{Introduction}

Software engineers reuse code to reduce development and maintenance costs. A
popular form of software reuse is the employment of Open-Source Software (OSS)
libraries, hosted on centralized code repositories, such as {\sc
Maven}\footnote{\href{https://search.maven.org/}{https://search.maven.org/}} or
{\sc \npm}.\footnote{\href{https://www.npmjs.com/}{https://www.npmjs.com/}} In
such settings, developers specify \emph{dependencies} to external library
releases in a textual file, that is then committed to the repository of the
\emph{client program}. Automated programs, usually referred to as \emph{package
managers}, resolve the dependency descriptions and connect to the central
repositories to download the specific library releases that are required to
build the client program.
Critically, dependencies can have dependencies of their own (\emph{transitive
dependencies}), and thus package managers need to resolve the transitive closure
of the dependency graph in order to build client applications.
Dependency versions are usually declared in a hierarchical versioning format,
which is, in most, but not all cases, a derivative of \emph{semantic
versioning}~\cite{semver}.
The co-evolving network of dependencies and end-user
applications is referred to as an \emph{ecosystem}.

The convenience of declarative specification for package reuse has led to
the massive adoption of package managers and package repositories.  While the
startup costs seem low for projects, reuse does not come for free. The software
engineering literature has thoroughly documented the hidden maintenance costs
around dependency reuse.  From the perspective of a library user, it is hard to
keep track of dependency updates especially for transitive
dependencies~\cite{DBLP:journals/corr/abs-1709-04621}, and assess their impact
on the client code base~\cite{bogart2016break}; the semantic versioning API
evolution provisions are rarely respected in practice by library
maintainers~\cite{raemaekers2017semantic}; and entrusting precious data on code
that the package manager automatically downloads is often not a wise
choice~\cite{DBLP:journals/corr/abs-1709-04621}. On the other hand, from the perspective of the
library maintainer, it is difficult to evolve APIs, for example by removing
methods~\cite{sawant2018reaction}, without breaking
clients~\cite{bogart2016break}, while incentives for professional maintenance of
library code are lacking~\cite{overney2020not}. In addition, as libraries may
also depend on other libraries, the library maintainers face the same issues as
library users face. Finally, the very nature of dependency networks and
the uncoordinated nature of their evolution burdens library users; recent
studies have shown that the average Javascript program has an estimated
mean of 80 transitive dependencies~\cite{zimmermann2019small} (up from 54 in
2017~\cite{DBLP:conf/msr/KikasGDP17}), while 50\% of the dependency sets change every 6 months in the Rust ecosystem~\cite{Hejderup18}.

As a result, in recent years, we have witnessed several spectacular failures
of package management networks, with severe implications on client programs, end users and the further adoption of OSS:

\begin{compactenum}

\item A dispute over a library name in the \npm\ ecosystem led to the removal
  of a library called~\texttt{leftpad}, which consisted of just 11 lines of
  code. The package removal led to the collapse of thousands of libraries which
  directly or transitively depended on \texttt{leftpad}, and hence a major
  disruption for client programs.

\item A company named Equifax leaked over 100.000 credit card records
    due to a dependency that was not updated. The compromised systems included
    a vulnerable version of the Apache Struts library, whose update was
    postponed as the Equifax security team erroneously underestimated the impact
    of the bug on their codebase.

\item Malicious developers uploaded to the Python package manager repository
  (PyPI) libraries whose name was deliberately misspelled, being almost
  identical to the original libraries (e.g., \texttt{urllib} instead of
  \texttt{urllib3}). The intention was to steal information from client
  applications of developers who had accidentally mistyped the library name in
  the dependency file.

\end{compactenum}

In this work, we argue that the main reason for such failures is that current
tools work at the wrong level of abstraction: while developers release and keep
track of \emph{packages}, actual reuse happens at the \emph{code} level. What we
propose is to use \emph{fine-grained network analysis}, as a more precise
instrument for the analysis of package ecosystems. We can obtain precise,
code-based reuse information by statically or dynamically analyzing the code at
the function/method call level. Static analysis comprises a set of
well-established techniques in the field of program analysis. While it
can derive sound(y)~\cite{livshits2015defense}
code reuse relationships, its precision is obstructed by the
dynamic nature of modern programming languages. On the other hand, dynamic
analysis can derive fully precise call relationships by instrumenting programs
while they are running, but its soundness
is limited by factors such as the availability of extensive test suites
or representative workloads.
Nonetheless, over-imposing call relationships on top of a package dependency network can lead
to more precise analyses, effectively augmenting the soundness properties
of the package dependency network with precision.

\section{Forges, product and dependencies}

At the dawn of the computer era, software development was mainly a solitary heroic activity of single men facing the complexity of problems with their bare hands in the darkness of
their caves; but those days have gone: modern software development relies more and more on existing third-party libraries, giving programmers the freedom to
focus on the core of what they have to do, delegating tedious or routine chores to reliable, specialized libraries they can
download from the Internet.

In a way, this is just the industrial revolution arriving at the harbor of software production.
In the words of Immanuel Kant: ``All trades, arts, and handiworks have gained by division of labour, namely, when, instead of one man doing everything,
each confines himself to a certain kind of work distinct from others in the treatment it requires, so as to be able to perform it with greater facility and
in the greatest perfection. Where the different kinds of work are not distinguished and divided, where everyone is a jack-of-all-trades, there manufactures
remain still in the greatest barbarism.''~\cite{Kant2002-KANGFT}

In free and open-source software, people share their work in the form of libraries, hosted on centralized code repositories, such as Maven\footnote{\url{https://search.maven.org/}}, \npm, RubyGems,
but also GitHub\footnote{\url{https://github.com/}} or SourceForge\footnote{\url{https://sourceforge.net/}}; some of these repositories are language-specific,
whereas others are not. We will broadly refer to such repositories as \emph{forges}.
Forges use different approaches in organizing the libraries or projects they host, and in fact they refer to such libraries in a variety of ways:
for example, \npm\ uses a flat organization of ``packages'', and the same does RubyGems with what they call ``gems'', whereas Maven organizes what they call ``artifacts'' into groups.
To avoid ambiguity, we shall prefer the abstract name \emph{product} to refer to all of them: a product is a coherent piece of software that can be used alone or as a library to develop something else.
We let $\Product$ be the set of all products\footnote{This set continuously evolves (typically, expands) in time; this fact will be ignored in this paper, as if we are taking a single
snapshot of the state of things at a certain moment in time.}, and we shall typically refer to a product $p \in \Product$ with a name, such as \texttt{org.apache.maven.plugins} or \texttt{org.slf4j}.

\smallskip
Every product in every forge exists typically in a number of \emph{revisions}: a new revision of a product is published to correct bugs or vulnerabilities found in previous releases,
or to introduce new functionalities. Every revision is identified in some way (e.g., by a version number or by a hashcode); the granularity (hence, the frequency) of releases changes from one repository to another (for example, in GitHub releases may actually be identified with commits
and are extremely frequent). We use the term \emph{version} to refer to the identifier (whatever it is) that specifies what revision of a specific product we are talking about
(e.g. \texttt{1.0.1-RC1}).
We let $\Revision$ be the set of all revisions, and we shall typically refer to a revision $r \in \Revision$ with a name combining the product name and the version,
such as \texttt{org.apache.maven.plugins-34} or \texttt{org.slf4j-2.0.0-alpha1}.
We also let
\[
\product: \Revision \to \Product
\]
be the function that returns the product corresponding to a given revision. For instance,
\[
	\product(\texttt{org.slf4j-2.0.0-alpha1})=\texttt{org.slf4j}
\]

\medskip
Any given revision (i.e., version of a product on a forge) is available in the form of a number of \emph{artifacts} (e.g., \text{jar} files, textual documents, zipped archives, etc.). Some of them contain the actual source code
of the library, whereas others contain metadata of various kind (makefiles, installation instructions, documentation, etc.). For the purposes of the current paper, we only consider the source code and
the so-called \emph{dependency specification}.
The latter metadata contains a description of the other products that are needed for this product to be compiled and/or executed: the syntax used in dependency specification depends on the forge.
The dependency specification of revision $r \in \Revision$ is denoted by $\dep(r)$.

A number of tools called \emph{package managers} are available that allow developers to specify which products their code depends on.
For maximum flexibility\footnote{Not all package managers allow for this level of flexibility. In most cases, only
one-element clauses are accepted.}, the dependency specification is expressed by a CNF logical formula~\cite{mancinelli2006managing}; more precisely,
the dependency specification is a set (interpreted as a logical conjunction) of \emph{dependency clauses}, where each
clause is itself a set (interpreted as a logical disjunction) of \emph{dependencies}.
A dependency is interpreted as a set of revisions of a product.

Here is an example of how a dependency is defined in Maven:
\begin{GrayBox}
\footnotesize
\begin{verbatim}
<plugin>
    <groupId>org.apache.maven.plugins</groupId>
    <artifactId>maven-compiler-plugin</artifactId>
    <version>34</version>
</plugin>
\end{verbatim}
\end{GrayBox}
and here is another one for Ivy:
\begin{GrayBox}
\footnotesize
\begin{verbatim}
<dependency org="org.slf4j"
    name="slf4j-api" rev="[1.7,)"/>
\end{verbatim}
\end{GrayBox}
In these examples, you see that dependencies can point to a \emph{specific} revision (in the first case, we are asking for version \texttt{34} of \texttt{org.apache.maven.plugins})
or to a set of revisions (in the second case, anything starting from version \texttt{1.7} of the \texttt{org.slf4j} library will fit).

\section{Package managers and dependency resolution}

Not only \emph{package managers} allow to specify which products a piece of code depends on, but also they automatize the process of downloading
recursively the dependencies of a given project and using them to build it. In order to describe how package managers work let us first provide some
definitions.

The \emph{global source dependency graph} is a directed graph whose node set is $\Revision$ and with an arc from $r$ to $r'$ whenever
$r'$ satisfies at least one of the dependencies in one of the clauses of $r$. The out-neighbors of a revision $r$ in the global source dependency graph
are called the (direct) dependants of $r$.

\begin{figure}
    \centering
    \includegraphics[scale=.4]{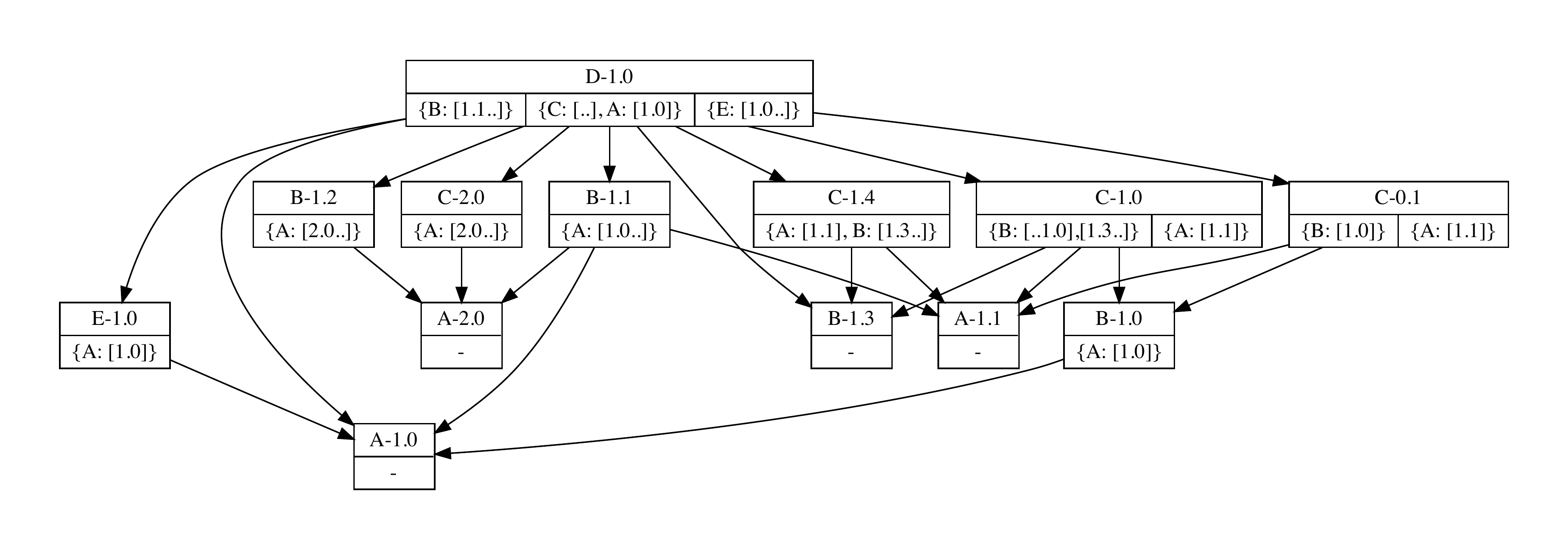}
    \caption{An example: the source dependency graph of revision \texttt{D-1.0}. In this example, we have only 13 revisions of five products (\texttt{A}, \texttt{B}, \texttt{C}, \texttt{D}, \texttt{E}).
    Each node is a revision: in the upper part
    we write the name of the revision (product and version); in the lower part its dependency specification: each rectangle is a dependency clause, represented as a set
    of dependencies. For example, revision \texttt{C-1.0} depends on revision \texttt{1.1} of \texttt{A}
    and on any revision of \texttt{B} with version $\leq$\texttt{1.0} or $\geq$\texttt{1.3}. As another example, \texttt{C-1.4} requires either revision \texttt{1.1} of \texttt{A}
    or any revision of \texttt{B} with version $\geq$\texttt{1.3}. Finally, \texttt{D-1.0} depends on revision $\geq$\texttt{1.1} of \texttt{B}, on any revision of \texttt{E} with version $\geq$\texttt{1.0} and then either
    on any revision of \texttt{C} or revision \texttt{1.0} of \texttt{A}.}
    \label{fig:dep}
\end{figure}

The \emph{source dependency graph} of revision $r_0 \in \Revision$
is the smallest subgraph of the global source dependency graph that includes $r_0$ and all the revisions that are reachable from $r_0$ in the
global source dependency graph. We show an example of source dependency graph in Figure~\ref{fig:dep}.

A given set of revisions $R \subseteq \Revision$ is said to satisfy a given dependency if it includes at least one element
satisfying the dependency; $R$ satisfies a given dependency clause if it satisfies at least one element of the clause; $R$
satisfies a given dependency specification if it satisfies all of its dependency clauses.

We say that $R$ is \emph{dependency-closed} iff it satisfies $\dep(r)$ for all $r \in R$. It is easy to see that the set of nodes of the source
dependency graph of $r_0$ is dependency-closed, but not necessarily the smallest dependency-closed set of revisions including $r_0$. Moreover,
it includes several revisions of the same product, which is in general undesirable.

\medskip
A package manager, given a revision $r_0\in\Revision$, finds a subset $R \subseteq\Revision$ that satisfies the following properties:

\begin{itemize}
  \item $r_0 \in R$
  \item $R$ is dependency-closed
  \item it does not contain different revisions of the same product; that is, if $r, r' \in R$ and $\product(r)=\product(r')$ then $r=r'$
  \item no proper subset of $R$ satisfies the above three conditions.
\end{itemize}

\begin{figure}[htbp]
    \centering
    \includegraphics[scale=.4]{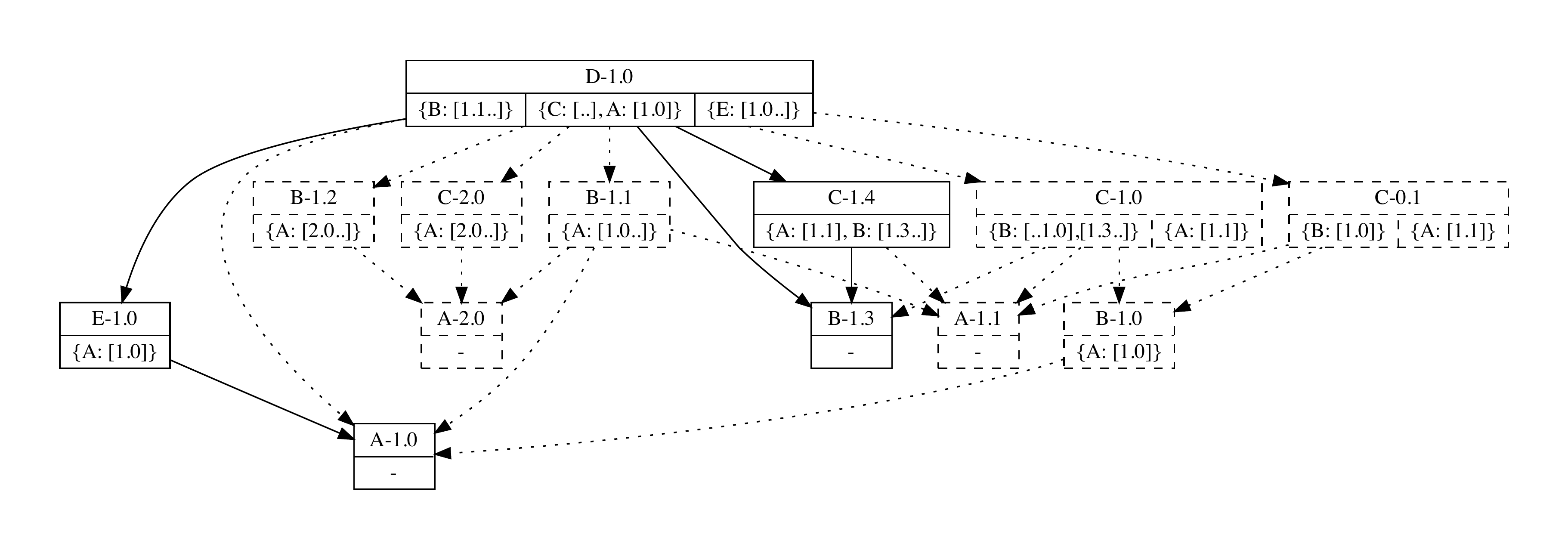}
    \caption{A resolved dependency graph of \texttt{D-1.0} obtained from the source dependency graph of Figure~\ref{fig:dep}. The dotted revisions are not included in the resolved
    dependency graph. As one can easily see, all dependencies are satisfied: for instance, \texttt{D-1.0} requires a version of \texttt{B} that is \text{1.1} or newer
    (and \texttt{B-1.3} is such), a version of \texttt{E} that is \texttt{1.0} or newer (\texttt{E-1.0} is such) and then either a revision of \texttt{C} or version \texttt{1.0} of \texttt{A} (and the former is provided, because the
    graph includes \texttt{C-1.4}).}
    \label{fig:resolved}
\end{figure}

In other words, the package manager should cherry-pick a subgraph of the source dependency graph of $r_0$ so that no more than one revision
of the same product is chosen but at the same time the resulting set is dependency-closed\footnote{It should be noted that real-world package managers
are more complex than this; we refer the interested reader to~\cite{ACGZ20} for details. Nonetheless, we think that the definition we are
using contains the core of what a package manager is supposed to achieve.}. This subgraph is called \emph{resolved dependency graph} of $r_0$.
Figure~\ref{fig:resolved} shows a possible resolved dependency graph for the example of Figure~\ref{fig:dep}.

\begin{figure}[hbtp]
    \centering
    \includegraphics[scale=.4]{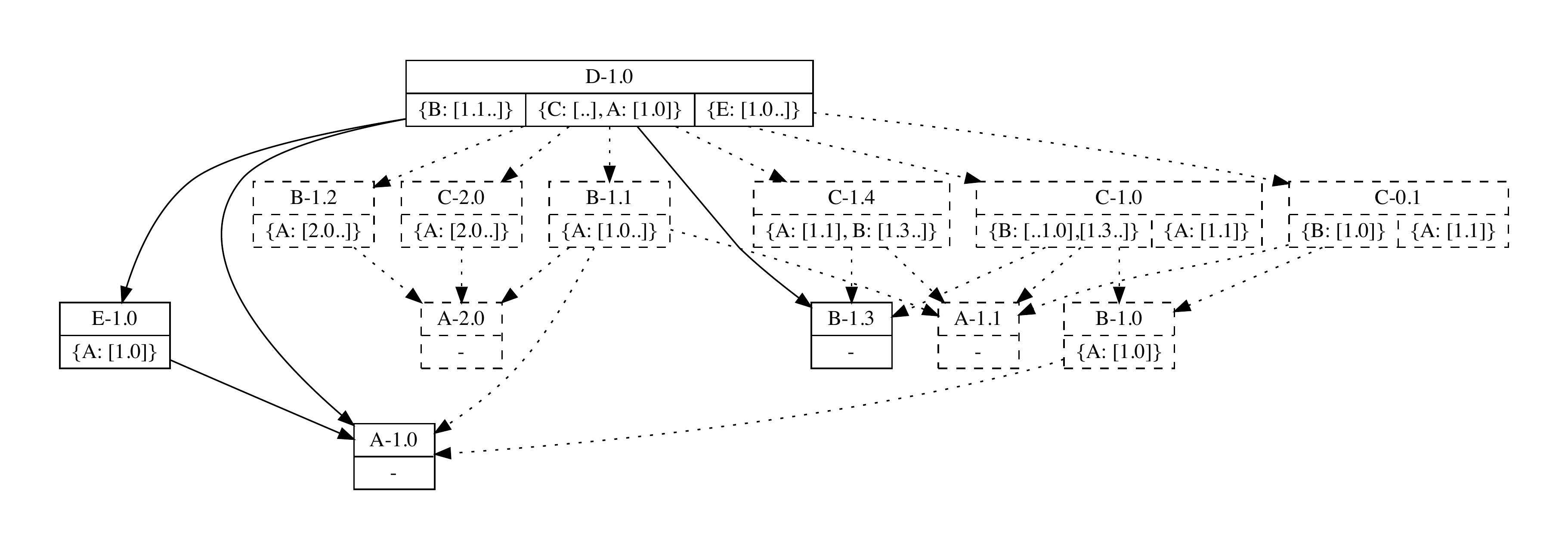}
    \caption{An alternative resolved dependency graph of \texttt{D-1.0} obtained from the source dependency graph of Figure~\ref{fig:dep}. Once more,
    all dependencies are satisfied: for instance \texttt{D-1.0} requires a version of \texttt{B} that is \texttt{1.1} or newer
    (and \texttt{B-1.3} is such) and then either a revision of \texttt{C} or version \texttt{1.0} of \texttt{A} (and the latter is provided because the
    graph includes \texttt{A-1.0}).}
    \label{fig:resolved-bis}
\end{figure}

Resolution is a process that can produce different (incomparable) sets of revisions: for example, Figure~\ref{fig:resolved-bis} is
an alternative resolved dependency graph for the same revision. Not only, different package managers use different resolution strategies
and may thus end up with different resolved dependency graphs, but even the same package manager may produce different resolutions
depending on some contextual information (e.g., a timestamp).

In the following, we assume that we have a single package manager and let $\resolved(r_0, c)\subseteq \Revision$ be the (node set of the) resolved dependency graph of $r_0$
obtained by the package manager in the context $c$.

\section{The price of reuse}

Public software forges, package managers and the resulting ecosystems made the dream of code reuse a reality, but this reality comes at a price.
These ecosystems are extremely fragile:
according to~\cite{DBLP:conf/msr/KikasGDP17}, in 2017 JavaScript products used to have an average of 54.6
products they depended upon (directly or transitively), with a steady growth of more than 60\% every year;
there are products in RubyGems that are in the transitive dependency of more than $400\,000$ other products
(meaning that if you remove one them, about 40\% of all the products in the ecosystem will cease to work).

Package \emph{users} gain great
value from reusing code, but they need to invest significant resources into
shielding themselves from software security, legal compliance and source code
incompatibility issues.

According to Snyk's annual 2019 report on the state of open-source security\footnote{\url{https://bit.ly/SoOSS2019}},
the number of security issues found in software almost doubles every two years (+44\% every year), and about 78\% of them are found in
indirect dependencies. This observation hints at how complicated sofware maintenance actually is: when a new vulnerability alert is found, for example,
it is essentially impossible to know whether the issue impacts on a specific product. Of course, the dependency graph can be used to
know if the impact is possible, but it is not enough to know if the specific piece of code that was broken is ever actually invoked (directly or indirectly) in
the product we are looking at, and in the positive case what are the functions that are put at risk and how the problem can be circumvented.

Even worse: 69\% of the developers are totally unaware of vulnerabilities existing in the products they depend upon, and 81.5\% of the systems
simply don't update their dependencies~\cite{DBLP:journals/corr/abs-1709-04621}.
This is probably because on one hand few tools are available to warn those developers in an automated way;
it is true, for example, that GitHub has recently launched an automated service
notifying repository owners
that they depend on packages affected by known security vulnerabilities,
but even so, it is like crying wolf: in most cases, vulnerabilities found in dependencies would not affect their software anyway.

\begin{figure}[htbp]
\centerline{\includegraphics[scale=0.6]{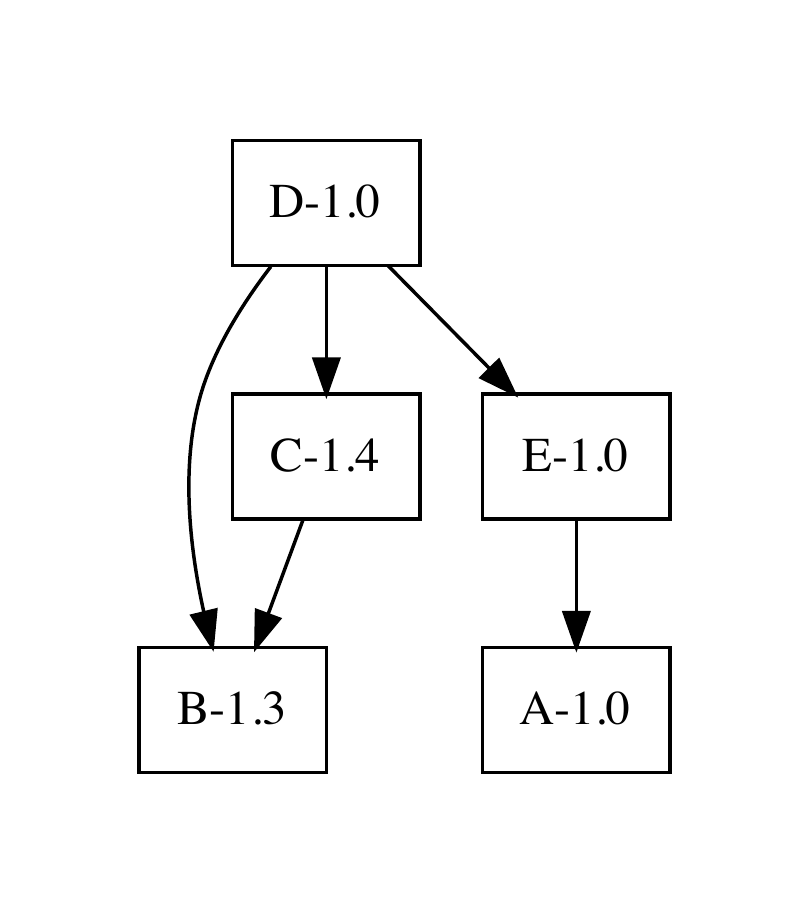}}
\caption{A fragment of the global source dependency graph, corresponding to the resolution of Figure~\ref{fig:resolved}.}
\label{fig:deps}
\end{figure}

\smallskip
As a concrete example, consider in Figure~\ref{fig:deps} the resolved dependency graph (the same as in Figure~\ref{fig:resolved} but where
we dropped the products that were not used in the resolution). This is in fact a fragment of the source dependency graph of Figure~\ref{fig:dep}.

\begin{figure}[htbp]
\centerline{\includegraphics[scale=0.6]{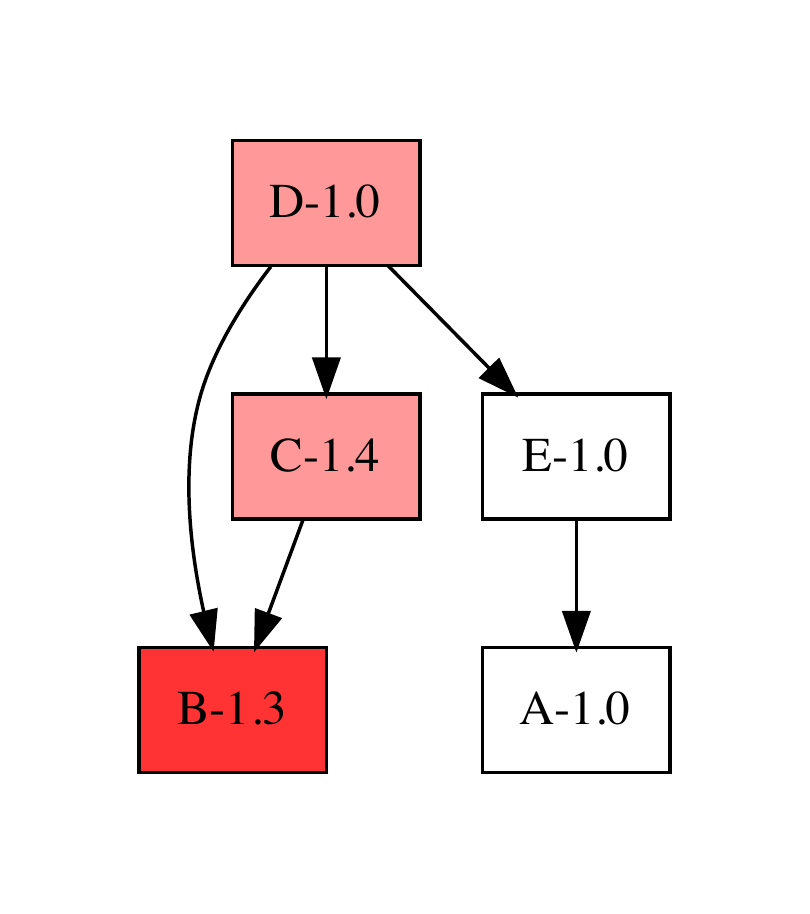}}
\caption{If a vulnerability is found in \texttt{B-1.3} (red in the picture), some other revisions in this picture may be at risk (light red in the picture).}
\label{fig:deps-infected}
\end{figure}

Suppose that a security alert is issued about revision \texttt{B-1.3}; then
by transitivity two other products involved are potentially infected (as shown in Figure~\ref{fig:deps-infected}). But is this really the case?

\medskip
From a different viewpoint, also legal and licensing issues are made quite complex by dependencies.
The complexity of licensing contracts and their effects on dependencies
is often underestimated by software developers. While several companies offer license
compliance checking services (e.g., BlackDuck
software, WhiteSource, FOSSA), a project's source code cannot be checked in isolation from its
dependency graph, and a project's dependency graph can extend to more components
than what specified in the package manager (e.g., implicit dependencies
on system libraries).

\smallskip
On the other hand, package \emph{providers} have no
reasonable means of evolving what they offer in a systematic way, because they are not
sure of the impact a change in their products, or in their licensing, can have on their clients.

The issues we just highlighted are related to the relatively naive design of package managers: they only resolve dependencies based on package
versions. As such, they cannot assess the risk of using dependencies,
they cannot notify developers of critical (e.g., security) updates, nor
can they assist them to evaluate the ecosystem impact of API evolution
tasks (e.g., removing a deprecated method).
Even if they were able to implement such functionalities, they could do so only at the bulk level of products,
but not at the level of function/method.


\begin{figure}
\centerline{\includegraphics[width=0.9\textwidth]{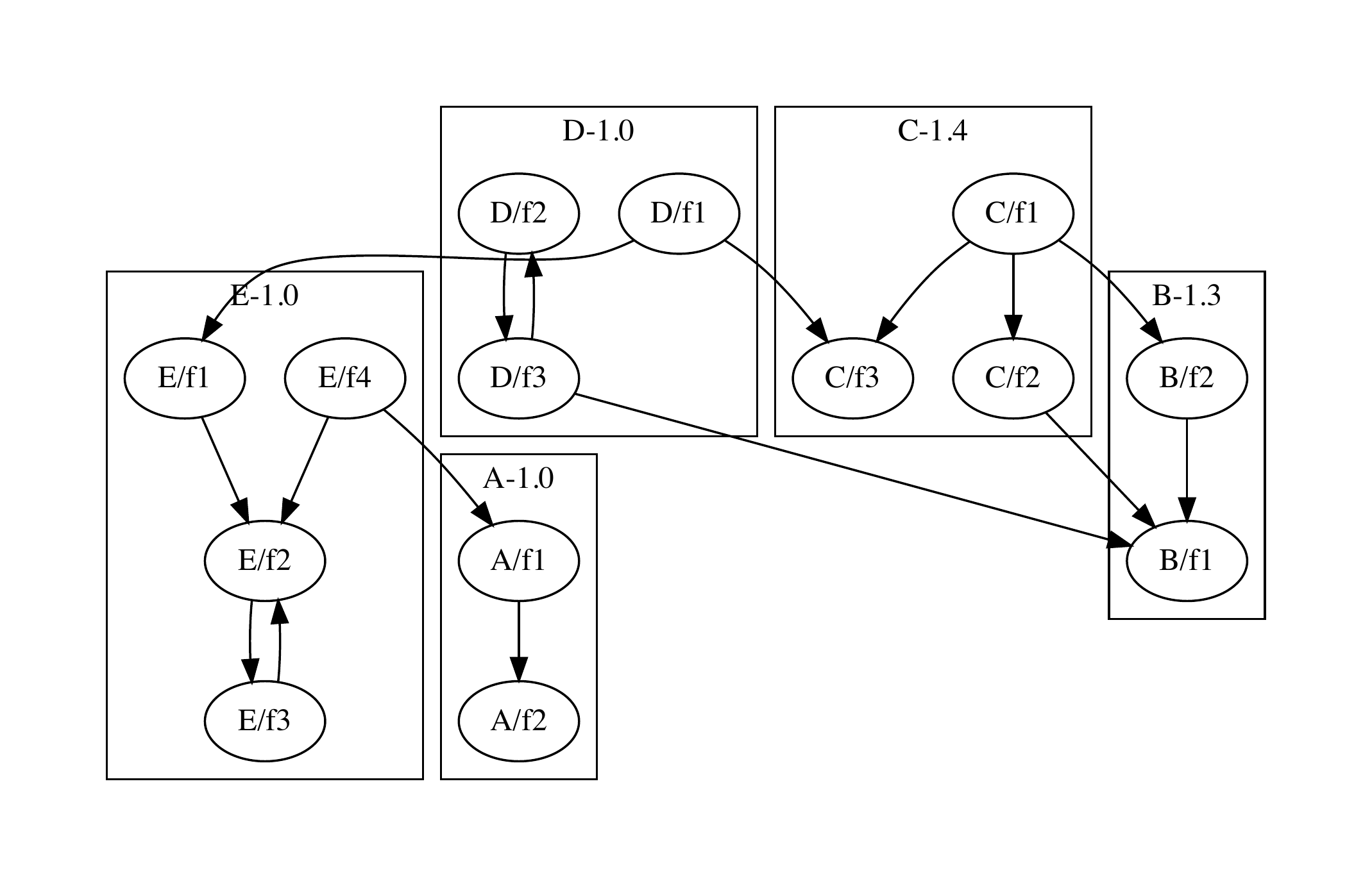}}
\caption{A function-level view of Figure~\ref{fig:deps}.}
\label{fig:deps-fine}
\end{figure}

\section{Network analysis as a\\software engineering tool}

While package-based dependency networks capture dependencies at the product
level, actual software reuse happens at the code level.
For example, Figure~\ref{fig:deps-fine} shows the same scenario depicted in
Figure~\ref{fig:deps}, but this time we can see the functions within each
revision, along with their calls. What we can observe is that a call from
\texttt{E/f4} to \texttt{A/f1} does not reach \texttt{A/f2}. However, \texttt{A/f2} is reachable from C/f3. In this
setting, a potential vulnerability in \texttt{A/f2} would render \texttt{C}
potentially vulnerable as well, whereas it would not affect \texttt{E}.
If we were able to identify exact function calling relationships between packages,
our whole analysis would become more precise.

Fortunately, the field of program analysis has been working for long to
automatically extract calling relationships in source code, in the form of call
graphs. Call graphs can be generated either \emph{statically}, by analyzing the
source code, or \emph{dynamically}, by instrumenting the code and tracing
program executions.  Static call graphs are neither complete (e.g., because they
miss calls by reflection or by dynamic dispatch), nor sound (some identified
execution paths may never materialize in real executions). Dynamic call graphs,
on the other hand, are strongly dependent on the test cases used to generate the
traces. Despite those shortcomings, call graphs are used in a variety
of use cases, notably, change impact analysis~\cite{ren2004chianti} and dead code
elimination~\cite{knoop1994partial}.

What we propose is to represent dependency relationships with call graphs.
Concretely, every (revision of a) product can be seen as a set of functions,
each calling other functions either belonging to the same product or to some of its dependants: it can be
abstractly represented as a directed graph with two types of nodes, internal and external. While internal nodes represent an actual function within
the same revision, external nodes are somehow less precisely identified --- they represent a function in some revision of some other product, but which
revision is not known, because it depends on the dependency-resolution process.

The whole dependency-resolution process depends on the choice of a resolution strategy adopted by the package manager, whose behavior is determined on
the specific revision of a specific product, we aim at using as a starting point. At the end of this process, we can actually identify
external nodes of each single revision involved with internal nodes of other revisions, obtaining the actual global call graph.

\section{(Stitched) call graphs}
Let us formalize the notion of call graph we have outlined. Every revision $r \in \Revision$ is associated with
a directed graph $G_r=(V_r, E_r)$, the \emph{call graph of $r$}, whose node set $V_r$ is bipartite $V_r=\vint_r \cup \vext_r$
into a set of internal nodes ($\vint_r$) and a set of external nodes ($\vext_r$); the latter nodes have no outgoing arcs.
Hence $E_r \subseteq \vint_r \times V_r$ (i.e., all arcs start from an internal node); the arcs themselves can be classified as being
internal (if they end up into an internal node) or external (if they end up into an external node).

External nodes (i.e., functions that exist in some dependant) carry sufficient metadata to allow one to identify them with some internal node of a dependant
regardless of how the resolution process was performed.
Formally, we can say that there is a function
\[
	\sigma_r: \vext_r \to 2^{\bigcup \vint_{r'}}
\]
where $2^X$ denotes the set of subsets of $X$, and the union ranges over all the out-neighbors $r'$ of $r$ in the global source dependency graph.
This function serves the purpose of mapping external nodes (i.e., calls to some library function) to internal nodes of dependants.

\begin{figure}
\centerline{\includegraphics[scale=.5]{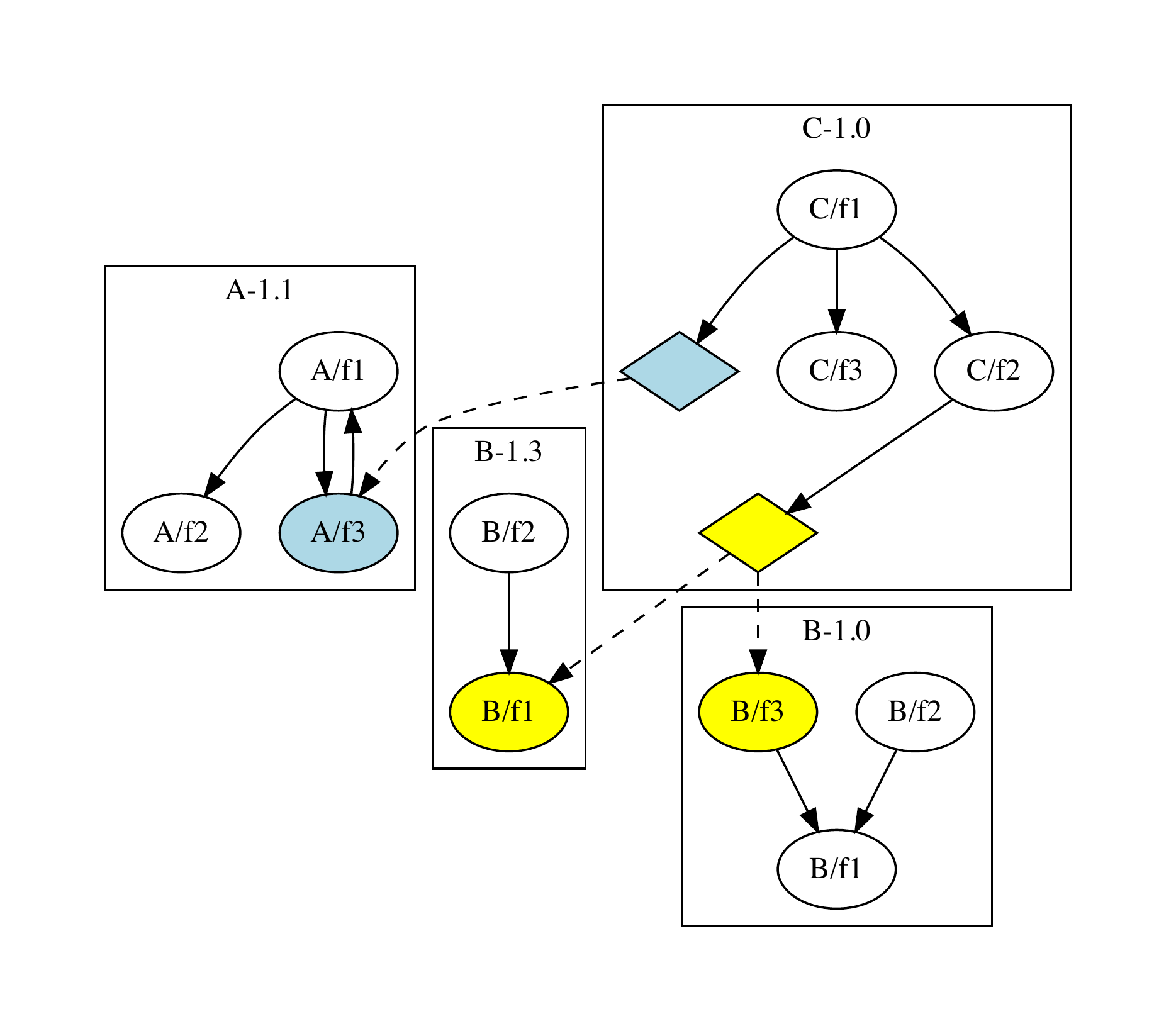}}
\caption{This picture shows the call graphs of \texttt{C-1.0} and its dependants (from Figure~\ref{fig:dep}). External nodes are diamond-shaped, and the dotted
arrows represent $\sigma$.}
\label{fig:deps-fine-sigma}
\end{figure}

As an example, Figure~\ref{fig:deps-fine-sigma} shows the call graphs of \texttt{C-1.0} and its dependants (from Figure~\ref{fig:dep}); external nodes are
diamond-shaped. The dotted arrows represent $\sigma$. We can see that \texttt{C-1.0} contains two external calls: \texttt{C/f1} and \texttt{C/f2} both call some external
function. Remember that, according to  Figure~\ref{fig:dep}, \texttt{C-1.0} depends on \texttt{A-1.1} and on any version of B excluding those
larger than \texttt{1.0} and smaller than \texttt{1.3}: we have only two revisions of \texttt{B} satisfying the constraint.
The yellow external nodes is mapped to the two internal nodes \texttt{B/f1} (of revision \texttt{B-1.3}) and \texttt{B/f3} (of revision \texttt{B-1.0}): which one will be used depends
on whether the resolution process chooses \texttt{B-1.3} or \texttt{B-1.0}. The blue external node corresponds to a call to a function of product \texttt{A}, and here
only \texttt{A-1.1} matches, with \texttt{A/f3} corresponding to the external call.

Using this function, we can define the notion of \emph{stitched call graph} of revision $r_0$ and a context $c$, written $G(r_0, c)$:
consider the graph obtained as a union of all $G_r$ with $r \in \resolved(r_0, c)$, and quotient its node set with respect to the smallest equivalence relation\footnote{Given
a graph $G=(V,E)$ and an equivalence relation $\approx$ between its nodes, the quotient graph $G/\approx$ has the set of equivalence classes $V/\approx$
as node set and an arc from $[x]$ to $[y]$ if and only if there is some $x' \approx x$ and some $y' \approx y$ such that $(x',y') \in E$.}
$x \sim y$ such that
$x \sim y$ whenever $y \in \sigma_r(x)$ for some  $r \in \resolved(r_0, c)$. This graph contains only internal nodes (because each external node is mapped to one or more
internal nodes by $\sigma_r$), and it is the function-level equivalent of $\resolved(r_0, c)$.

\begin{figure}
  \centerline{\includegraphics[scale=0.5]{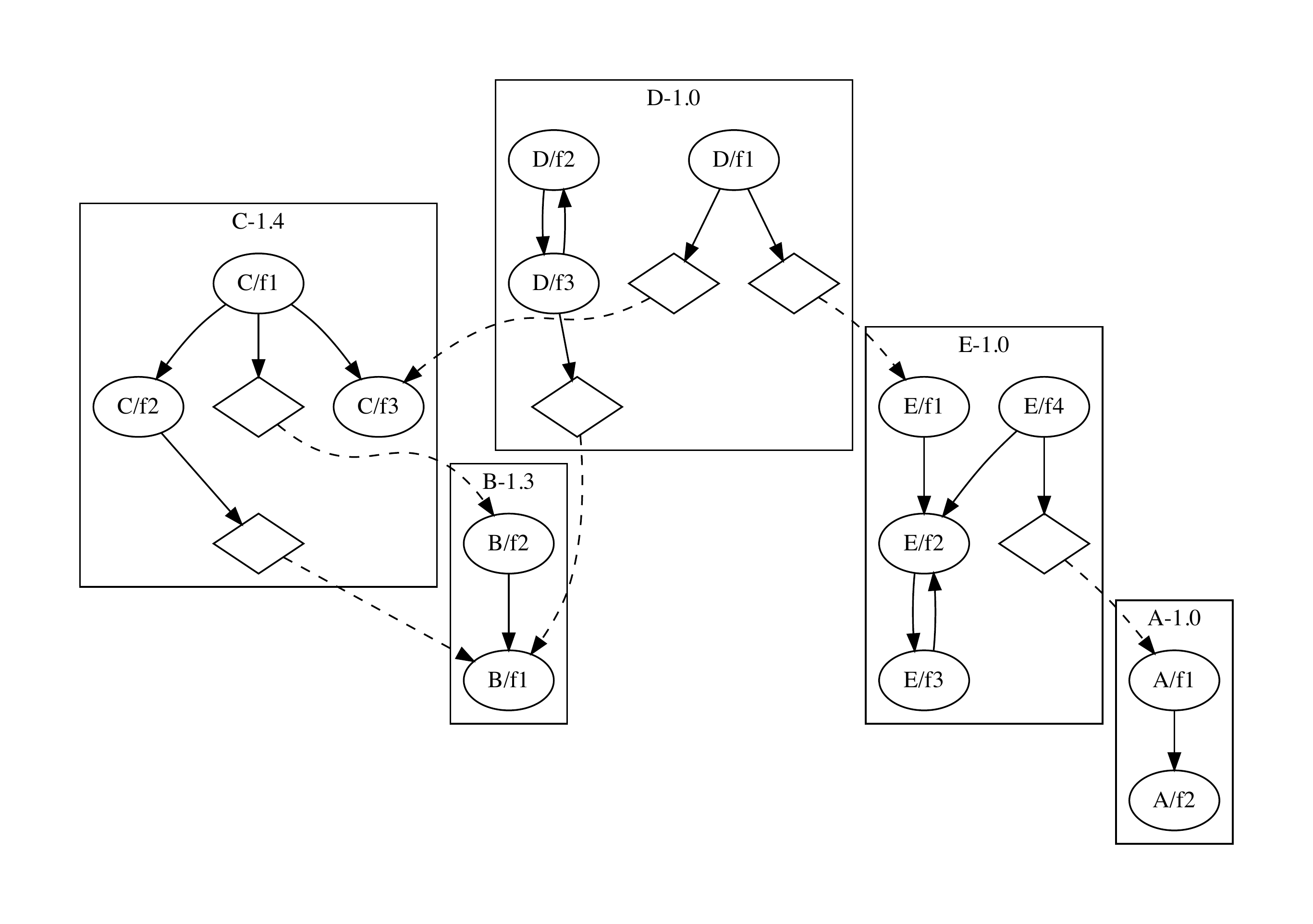}}
\caption{This picture shows the call graphs of the revisions involved in Figure~\ref{fig:resolved}: within each graph, external nodes are diamond-shaped.
The dotted arcs correspond to the map $\sigma$ (the function will map each external nodes to possibly many internal nodes in many dependants, but
here we are looking at a resolved graph). The graph obtained after stitching is that of Figure~\ref{fig:deps-fine}.}
\label{fig:deps-fine-cg}
\end{figure}

We call this graph ``stitched'' because it is obtained by stitching together the call graphs of a resolved dependency graph
identifying every external node with a set of internal nodes of dependants. Figure~\ref{fig:deps-fine-cg} shows how the stitched call graph construction
works.

\section{Opportunities \& Challenges}

\begin{figure}
\centerline{\includegraphics[scale=.5]{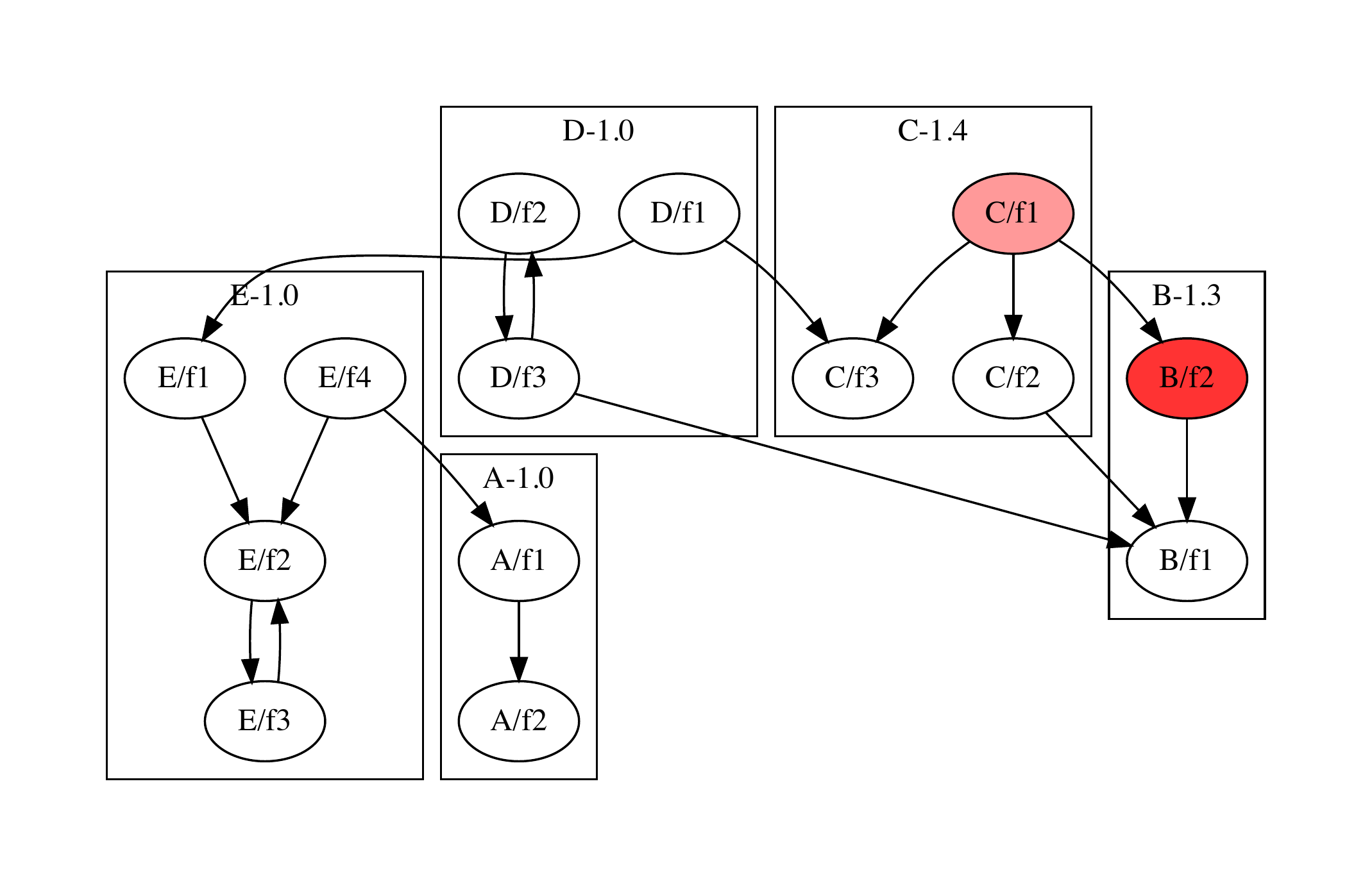}}
\caption{The function-level analysis shows that if the vulnerability in \texttt{B-1.3} is because of function \texttt{B/f2}, then only \texttt{C/f1} is at risk.
In particular, none of the functions within library \texttt{D-1.0} is involved.}
\label{fig:deps-fine-infected}
\end{figure}

We present examples on how the availability of stitched call graphs can improve
the reliability of software development practices, and improve the robustness of
the whole ecosystem:

\begin{itemize}
\item Every time a bug or a security breach is found in a library, developers
will be able to precisely analyze whether their applications are
calling into vulnerable code and decide whether dependency updates are
necessary; the ecosystem itself will be able to notify the developers of
vulnerable applications in real-time, after a security issue has been disclosed.
This type of functionality would have been beneficial in preventing the Equifax breach.
By analyzing security alerts at a function level, you can get
much more precise information and know exactly  which parts of your code need to be
fixed: Figure~\ref{fig:deps-fine-infected} shows that in our example revision \texttt{D-1.0} is not at all impacted by a bug found in
function \texttt{B/f2}. Similarly, only a portion of the functions of \texttt{C-1.4} are impacted.

\item Using the call graph, one can precisely identify the ecosystem-wide impact
  (direct and transitive) that any API change can have. This kind of
change impact analysis can thus become a first-class tool for software developers (much like a debugger or a profiler is).
Developers will be able to get quantitative answers to questions
such as ``\emph{How many packages are affected if I remove a certain
method/interface?}'' and will be able to make decisions and proactively notify downstream packages
for breaking changes when an upstream API has changed. The availability of this
functionality would have prevented the \texttt{left-pad} incident, for example.

\item The fact that licensing is usually only evaluated at the library level
and not at the function level introduces (at least conceptually) a rigidity that
is not desirable for organizations releasing open source software components.
For example, a library may include subroutines with
different licensing contracts. Using the call graph we can check that function-level licenses of our own software are consistent with one another, and that they are consistent with the
licenses attached to the libraries our software depends on.
\end{itemize}

More generally, the call graph contains a big deal of information about software that could not be obtained otherwise. In particular, the notion of centrality~\cite{BoVAC} applied to
the call graph can determine which parts of the software ecosystem are more relevant; this information can be used
to target critical or risky functions, or to better concentrate maintenance efforts of large
software repositories. This path can be thought of as moving one step beyond the traditional approach to profiling, using
network analysis methods as the weapon of choice.

Despite the obvious advantages, analyzing whole ecosystems at the function level
is not trivial. We foresee the following challenges:

\paragraph*{Scale}
The unified call graph Hejderup et al.~\cite{Hejderup18} built
for Rust contained 6 million nodes and 16 million edges. However, Rust as an
ecosystem featuring 250k revisions is \emph{an order of magnitude}
smaller than that of Java (2.5M) or Python (1.2M).
Even assuming a prudent estimate of about 100 functions per revision, we are talking of graphs with
about $10^{11}$ nodes, and thousands of billions of arcs!

These graphs are bound to challenge the current state-of-art in graph
processing systems, especially considering that those graphs change frequently and
that they materialize dynamically based on the dependency-resolution strategy.
In addition, such graphs will need to be queried (e.g.,
traversed and sliced) in real time by clients.

This challenge calls for new, aggressive, dynamic compression techniques, specially tailored around the
structure and topology of call graphs, that can go beyond the state-of-art in graph compression~\cite{DBLP:reference/bdt/BoldiV19}.
For deeper analysis and ranking it might be necessary to store in compact form some metadata about the actual calls. For
example, the users might contribute profiling data making it possible to
decorate arcs of the graph with the estimated number of times a particular
function is called at a specific location in the code: such information
would be invaluable in the ranking process, but it would
require further storage and new as-yet unknown compression techniques.

\paragraph*{Call graph soundness} Soundness is an important property of static analyses. Unsound analyses
lead to reporting false positives to developers (e.g., a piece of code is
labeled as having a bug, but in reality, it does not), which in turn
makes developers not trust the analyses.
Unfortunately, creating sound static analyses is known to be an undecidable problem (Rice's theorem).
For the proposed system to be successful, the right balance between soundness and
usefulness should be reached.

The problem of unsoundness translates to either under- or over-approximation
of a program's behavior.
In practical terms, and since nodes (functions) can always be fully recovered,
the ecosystem call graph may contain more or less edges than what we could
obtain from a program execution trace.
To mitigate this problem, we can employ two strategies: crowdsourcing the
call graph augmentation and machine learning on graphs.
In the first case, users of the system can be asked to instrument their
test runs or actual deployments with dynamic call graph extraction tools.
A centralized location keeping track of the call graphs can compare the uploaded
traces and create edges in case they are missing; full trust can be given to
such traces, as dynamic call graphs are \emph{de facto} precise.
Finally, network analysis can play an important role here. Specifically, one could
envisage methods that exploit the structural and evolutionary properties of the
graph to employ algorithms, such as friend recommendation, in order to augment or
prune the ecosystem call graph.

\paragraph*{Bringing value to developers}
What we have just described is only the backbone behind a set of tools and services that should integrate with
the final developer's programming environment (e.g., in the form of plugins for the programmers' favorite IDE) as well as with continuous integration
tools.
To make the call graphs available and to enable additional analyses (e.g.,
change impact analysis, security/bug propagation etc), the FASTEN
project~\footnote{\url{https://www.fasten-project.eu}} develops a continuously
updated service that will act as a hub for ecosystem-scale analysis for the Java,
Python, C and Rust programming communities.

\begin{figure}
\centering
\hspace*{-2.5cm}
\includegraphics[scale=.3]{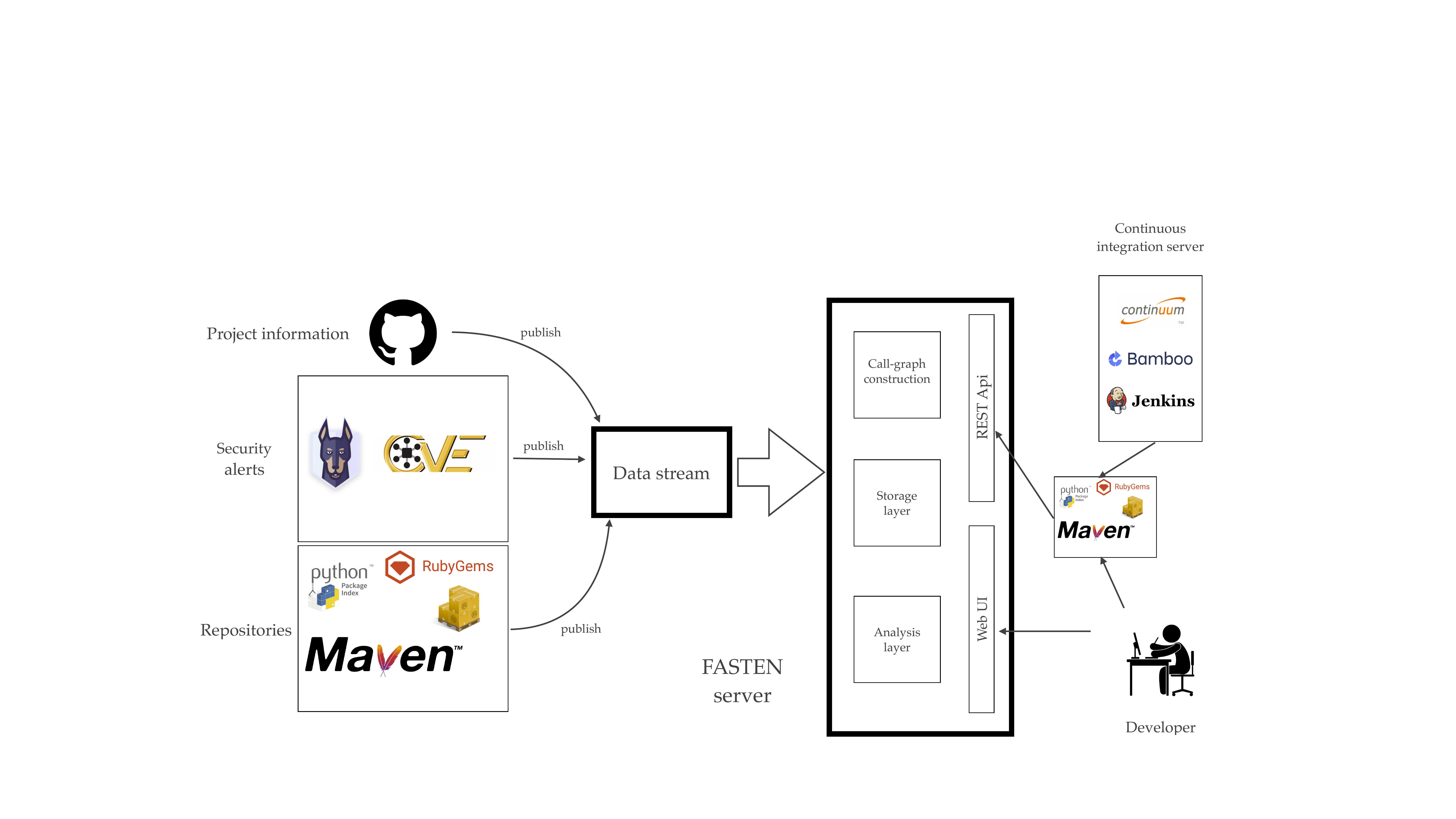}
\caption{\label{fig:flow}The FASTEN EU Project toolchain.}
\end{figure}

\smallskip
The ongoing FASTEN EU Project proposes an innovative approach to solve the above challenges; FASTEN design is based on the streaming toolchain sketched in Figure~\ref{fig:flow}: data
coming from different sources are streamed to a centralized server that extracts call graphs, stores them in compressed form and prepares efficient data structures (indices) to query them efficiently.
Front-end tools are provided for developers to interrogate the call-graph database.

The preliminary results, especially for what concerns call-graph compression and analysis, are very promising.
The first nontrivial problem related to call graphs is about their compressibility: call graphs are a relatively new object, and it is unclear whether standard compression
techniques~\cite{DBLP:reference/bdt/BoldiV19} can be fruitfully applied. It turns out that call graphs can be indeed compressed very well, for instance using
variants of LLP~\cite{DBLP:conf/www/BoldiRSV11}, a technique initially developed to extend webgraph compression techniques to more general types of graphs (e.g., social
networks).

A second important preliminary observation is that, despite their size, call graphs exhibit relatively short call
chains, allowing for efficient resolution of reachability and co-reachability queries~\cite{YuCGRQS}, which are the fundamental ingredient of most change-impact studies.

These early findings suggest that the challenges we envision can be won, leading to a new, efficient, safer and more productive
software-writing environment.

\section{Conclusions}

Since the introduction of software modularization~\cite{parnas1972criteria},
software engineers have long been trying to make software reusable.
Technologies such as object-oriented programming, components, commercial off-the-shelf libraries and aspects have all touted the reuse horn, to various degrees of success.
During the last 15 years, package managers and the open-source software movement have succeeded in making the dream of software reuse a tangible reality.
However, this reality is not without problems.
The current state of practice and the frequent, spectacular failures of modern
software ecosystems point towards the limits of what comprises the first generation of package management technology.

In this vision paper, we presented a new design for package
management systems that can, in a large degree, overcome the fallacies of
current ones and pave the way to new, exciting reuse possibilities.
What we need to do is change our unit of reuse from the package to the
function/method and embrace network analysis as first-class citizen in
future software engineering tools.

\section*{Acknowledgment}

We would like to express our gratitude to Sebastiano Vigna and Stefano
Zacchiroli for their precious help. This work has been partially funded by the FASTEN
EU Project, H2020-ICT-2018-2020 (Information and Communication Technologies).

\bibliographystyle{ACM-Reference-Format}
\bibliography{biblio,extra}


\hyphenation{ Vi-gna Sa-ba-di-ni Kath-ryn Ker-n-i-ghan Krom-mes Lar-ra-bee
  Pat-rick Port-able Post-Script Pren-tice Rich-ard Richt-er Ro-bert Sha-mos
  Spring-er The-o-dore Uz-ga-lis }
\begin{thebibliography}{20}


\ifx \showCODEN    \undefined \def \showCODEN     #1{\unskip}     \fi
\ifx \showDOI      \undefined \def \showDOI       #1{#1}\fi
\ifx \showISBNx    \undefined \def \showISBNx     #1{\unskip}     \fi
\ifx \showISBNxiii \undefined \def \showISBNxiii  #1{\unskip}     \fi
\ifx \showISSN     \undefined \def \showISSN      #1{\unskip}     \fi
\ifx \showLCCN     \undefined \def \showLCCN      #1{\unskip}     \fi
\ifx \shownote     \undefined \def \shownote      #1{#1}          \fi
\ifx \showarticletitle \undefined \def \showarticletitle #1{#1}   \fi
\ifx \showURL      \undefined \def \showURL       {\relax}        \fi
\providecommand\bibfield[2]{#2}
\providecommand\bibinfo[2]{#2}
\providecommand\natexlab[1]{#1}
\providecommand\showeprint[2][]{arXiv:#2}

\bibitem[\protect\citeauthoryear{{Abate}, {Di Cosmo}, {Gousios}, and
  {Zacchiroli}}{{Abate} et~al\mbox{.}}{2020}]%
        {ACGZ20}
\bibfield{author}{\bibinfo{person}{P. {Abate}}, \bibinfo{person}{R. {Di
  Cosmo}}, \bibinfo{person}{G. {Gousios}}, {and} \bibinfo{person}{S.
  {Zacchiroli}}.} \bibinfo{year}{2020}\natexlab{}.
\newblock \showarticletitle{Dependency Solving Is Still Hard, but We Are
  Getting Better at It}. In \bibinfo{booktitle}{\emph{The 27th IEEE
  International Conference on Software Analysis, Evolution and Reengineering
  (SANER)}} (London, Ontario, Canada). \bibinfo{pages}{547--551}.
\newblock
\urldef\tempurl%
\url{https://doi.org/10.1109/SANER48275.2020.9054837}
\showDOI{\tempurl}


\bibitem[\protect\citeauthoryear{Bogart, K{\"a}stner, Herbsleb, and
  Thung}{Bogart et~al\mbox{.}}{2016}]%
        {bogart2016break}
\bibfield{author}{\bibinfo{person}{Christopher Bogart},
  \bibinfo{person}{Christian K{\"a}stner}, \bibinfo{person}{James Herbsleb},
  {and} \bibinfo{person}{Ferdian Thung}.} \bibinfo{year}{2016}\natexlab{}.
\newblock \showarticletitle{How to break an API: cost negotiation and community
  values in three software ecosystems}. In
  \bibinfo{booktitle}{\emph{Proceedings of the 2016 24th ACM SIGSOFT
  International Symposium on Foundations of Software Engineering}}. ACM,
  \bibinfo{pages}{109--120}.
\newblock


\bibitem[\protect\citeauthoryear{Boldi, Rosa, Santini, and Vigna}{Boldi
  et~al\mbox{.}}{2011}]%
        {DBLP:conf/www/BoldiRSV11}
\bibfield{author}{\bibinfo{person}{Paolo Boldi}, \bibinfo{person}{Marco Rosa},
  \bibinfo{person}{Massimo Santini}, {and} \bibinfo{person}{Sebastiano Vigna}.}
  \bibinfo{year}{2011}\natexlab{}.
\newblock \showarticletitle{Layered label propagation: a multiresolution
  coordinate-free ordering for compressing social networks}. In
  \bibinfo{booktitle}{\emph{Proceedings of the 20th International Conference on
  World Wide Web, {WWW} 2011, Hyderabad, India, March 28 - April 1, 2011}},
  \bibfield{editor}{\bibinfo{person}{Sadagopan Srinivasan},
  \bibinfo{person}{Krithi Ramamritham}, \bibinfo{person}{Arun Kumar},
  \bibinfo{person}{M.~P. Ravindra}, \bibinfo{person}{Elisa Bertino}, {and}
  \bibinfo{person}{Ravi Kumar}} (Eds.). \bibinfo{publisher}{{ACM}},
  \bibinfo{pages}{587--596}.
\newblock
\urldef\tempurl%
\url{https://doi.org/10.1145/1963405.1963488}
\showDOI{\tempurl}


\bibitem[\protect\citeauthoryear{Boldi and Vigna}{Boldi and Vigna}{2014}]%
        {BoVAC}
\bibfield{author}{\bibinfo{person}{Paolo Boldi} {and}
  \bibinfo{person}{Sebastiano Vigna}.} \bibinfo{year}{2014}\natexlab{}.
\newblock \showarticletitle{Axioms for Centrality}.
\newblock \bibinfo{journal}{\emph{Internet Math.}} \bibinfo{volume}{10},
  \bibinfo{number}{3-4} (\bibinfo{year}{2014}), \bibinfo{pages}{222--262}.
\newblock


\bibitem[\protect\citeauthoryear{Boldi and Vigna}{Boldi and Vigna}{2019}]%
        {DBLP:reference/bdt/BoldiV19}
\bibfield{author}{\bibinfo{person}{Paolo Boldi} {and}
  \bibinfo{person}{Sebastiano Vigna}.} \bibinfo{year}{2019}\natexlab{}.
\newblock \showarticletitle{(Web/Social) Graph Compression}.
\newblock In \bibinfo{booktitle}{\emph{Encyclopedia of Big Data
  Technologies.}}, \bibfield{editor}{\bibinfo{person}{Sherif Sakr} {and}
  \bibinfo{person}{Albert~Y. Zomaya}} (Eds.). \bibinfo{publisher}{Springer}.
\newblock
\urldef\tempurl%
\url{https://doi.org/10.1007/978-3-319-63962-8\_54-1}
\showDOI{\tempurl}


\bibitem[\protect\citeauthoryear{Hejderup, Beller, and Gousios}{Hejderup
  et~al\mbox{.}}{2018}]%
        {Hejderup18}
\bibfield{author}{\bibinfo{person}{Joseph Hejderup}, \bibinfo{person}{Moritz
  Beller}, {and} \bibinfo{person}{Georgios Gousios}.}
  \bibinfo{year}{2018}\natexlab{}.
\newblock \bibinfo{booktitle}{\emph{Building a Unified Call Graph at Ecosystem
  Level}}.
\newblock \bibinfo{type}{{T}echnical {R}eport} TUD-SERG-2018-002.
  \bibinfo{institution}{Delft University of Techology}. \bibinfo{pages}{20}
  pages.
\newblock
\urldef\tempurl%
\url{http://gousios.org/pubs/ucg.pdf}
\showURL{%
\tempurl}
\newblock
\shownote{Online: \url{http://gousios.org/pub/ucg.pdf}.}


\bibitem[\protect\citeauthoryear{Kant}{Kant}{1785}]%
        {Kant2002-KANGFT}
\bibfield{author}{\bibinfo{person}{Immanuel Kant}.} \bibinfo{year}{2002
  [1785]}\natexlab{}.
\newblock \bibinfo{booktitle}{\emph{Groundwork for the Metaphysics of Morals}}.
\newblock \bibinfo{publisher}{Oxford University Press}.
\newblock


\bibitem[\protect\citeauthoryear{Kikas, Gousios, Dumas, and Pfahl}{Kikas
  et~al\mbox{.}}{2017}]%
        {DBLP:conf/msr/KikasGDP17}
\bibfield{author}{\bibinfo{person}{Riivo Kikas}, \bibinfo{person}{Georgios
  Gousios}, \bibinfo{person}{Marlon Dumas}, {and} \bibinfo{person}{Dietmar
  Pfahl}.} \bibinfo{year}{2017}\natexlab{}.
\newblock \showarticletitle{Structure and evolution of package dependency
  networks}. In \bibinfo{booktitle}{\emph{Proceedings of the 14th International
  Conference on Mining Software Repositories, {MSR} 2017, Buenos Aires,
  Argentina, May 20-28, 2017}}. \bibinfo{pages}{102--112}.
\newblock
\urldef\tempurl%
\url{https://doi.org/10.1109/MSR.2017.55}
\showDOI{\tempurl}


\bibitem[\protect\citeauthoryear{Knoop, R{\"u}thing, and Steffen}{Knoop
  et~al\mbox{.}}{1994}]%
        {knoop1994partial}
\bibfield{author}{\bibinfo{person}{Jens Knoop}, \bibinfo{person}{Oliver
  R{\"u}thing}, {and} \bibinfo{person}{Bernhard Steffen}.}
  \bibinfo{year}{1994}\natexlab{}.
\newblock \showarticletitle{Partial dead code elimination}.
\newblock \bibinfo{journal}{\emph{ACM SIGPLAN Notices}} \bibinfo{volume}{29},
  \bibinfo{number}{6} (\bibinfo{year}{1994}), \bibinfo{pages}{147--158}.
\newblock


\bibitem[\protect\citeauthoryear{Kula, Germ{\'{a}}n, Ouni, Ishio, and
  Inoue}{Kula et~al\mbox{.}}{2017}]%
        {DBLP:journals/corr/abs-1709-04621}
\bibfield{author}{\bibinfo{person}{Raula~Gaikovina Kula},
  \bibinfo{person}{Daniel~M. Germ{\'{a}}n}, \bibinfo{person}{Ali Ouni},
  \bibinfo{person}{Takashi Ishio}, {and} \bibinfo{person}{Katsuro Inoue}.}
  \bibinfo{year}{2017}\natexlab{}.
\newblock \showarticletitle{Do Developers Update Their Library Dependencies? An
  Empirical Study on the Impact of Security Advisories on Library Migration}.
\newblock \bibinfo{journal}{\emph{CoRR}}  \bibinfo{volume}{abs/1709.04621}
  (\bibinfo{year}{2017}).
\newblock
\showeprint[arxiv]{1709.04621}
\urldef\tempurl%
\url{http://arxiv.org/abs/1709.04621}
\showURL{%
\tempurl}


\bibitem[\protect\citeauthoryear{Livshits, Sridharan, Smaragdakis, Lhot{\'a}k,
  Amaral, Chang, Guyer, Khedker, M{\o}ller, and Vardoulakis}{Livshits
  et~al\mbox{.}}{2015}]%
        {livshits2015defense}
\bibfield{author}{\bibinfo{person}{Benjamin Livshits}, \bibinfo{person}{Manu
  Sridharan}, \bibinfo{person}{Yannis Smaragdakis},
  \bibinfo{person}{Ond{\v{r}}ej Lhot{\'a}k}, \bibinfo{person}{J~Nelson Amaral},
  \bibinfo{person}{Bor-Yuh~Evan Chang}, \bibinfo{person}{Samuel~Z Guyer},
  \bibinfo{person}{Uday~P Khedker}, \bibinfo{person}{Anders M{\o}ller}, {and}
  \bibinfo{person}{Dimitrios Vardoulakis}.} \bibinfo{year}{2015}\natexlab{}.
\newblock \showarticletitle{In defense of soundiness: A manifesto}.
\newblock \bibinfo{journal}{\emph{Commun. ACM}} \bibinfo{volume}{58},
  \bibinfo{number}{2} (\bibinfo{year}{2015}), \bibinfo{pages}{44--46}.
\newblock


\bibitem[\protect\citeauthoryear{Mancinelli, Boender, Di~Cosmo, Vouillon,
  Durak, Leroy, and Treinen}{Mancinelli et~al\mbox{.}}{2006}]%
        {mancinelli2006managing}
\bibfield{author}{\bibinfo{person}{Fabio Mancinelli}, \bibinfo{person}{Jaap
  Boender}, \bibinfo{person}{Roberto Di~Cosmo}, \bibinfo{person}{Jerome
  Vouillon}, \bibinfo{person}{Berke Durak}, \bibinfo{person}{Xavier Leroy},
  {and} \bibinfo{person}{Ralf Treinen}.} \bibinfo{year}{2006}\natexlab{}.
\newblock \showarticletitle{Managing the complexity of large free and open
  source package-based software distributions}. In
  \bibinfo{booktitle}{\emph{21st IEEE/ACM International Conference on Automated
  Software Engineering (ASE'06)}}. IEEE, \bibinfo{pages}{199--208}.
\newblock


\bibitem[\protect\citeauthoryear{Overney, Meinicke, K{\"a}stner, and
  Vasilescu}{Overney et~al\mbox{.}}{2020}]%
        {overney2020not}
\bibfield{author}{\bibinfo{person}{Cassandra Overney}, \bibinfo{person}{Jens
  Meinicke}, \bibinfo{person}{Christian K{\"a}stner}, {and}
  \bibinfo{person}{Bogdan Vasilescu}.} \bibinfo{year}{2020}\natexlab{}.
\newblock \showarticletitle{How to Not Get Rich: An Empirical Study of
  Donations in Open Source}. In \bibinfo{booktitle}{\emph{Proceedings of the
  2020 42th International Conference on Software Engineering}}.
\newblock


\bibitem[\protect\citeauthoryear{Parnas}{Parnas}{1972}]%
        {parnas1972criteria}
\bibfield{author}{\bibinfo{person}{David~L Parnas}.}
  \bibinfo{year}{1972}\natexlab{}.
\newblock \showarticletitle{On the criteria to be used in decomposing systems
  into modules}.
\newblock In \bibinfo{booktitle}{\emph{Pioneers and Their Contributions to
  Software Engineering}}. \bibinfo{publisher}{Springer},
  \bibinfo{pages}{479--498}.
\newblock


\bibitem[\protect\citeauthoryear{Preston-Werner}{Preston-Werner}{[n.d.]}]%
        {semver}
\bibfield{author}{\bibinfo{person}{Tom Preston-Werner}.}
  \bibinfo{year}{[n.d.]}\natexlab{}.
\newblock \bibinfo{title}{Semantic Versioning 2.0.0}.
\newblock
\newblock
\urldef\tempurl%
\url{https://semver.org}
\showURL{%
\tempurl}


\bibitem[\protect\citeauthoryear{Raemaekers, van Deursen, and
  Visser}{Raemaekers et~al\mbox{.}}{2017}]%
        {raemaekers2017semantic}
\bibfield{author}{\bibinfo{person}{Steven Raemaekers}, \bibinfo{person}{Arie
  van Deursen}, {and} \bibinfo{person}{Joost Visser}.}
  \bibinfo{year}{2017}\natexlab{}.
\newblock \showarticletitle{Semantic versioning and impact of breaking changes
  in the Maven repository}.
\newblock \bibinfo{journal}{\emph{Journal of Systems and Software}}
  \bibinfo{volume}{129} (\bibinfo{year}{2017}), \bibinfo{pages}{140--158}.
\newblock


\bibitem[\protect\citeauthoryear{Ren, Shah, Tip, Ryder, and Chesley}{Ren
  et~al\mbox{.}}{2004}]%
        {ren2004chianti}
\bibfield{author}{\bibinfo{person}{Xiaoxia Ren}, \bibinfo{person}{Fenil Shah},
  \bibinfo{person}{Frank Tip}, \bibinfo{person}{Barbara~G Ryder}, {and}
  \bibinfo{person}{Ophelia Chesley}.} \bibinfo{year}{2004}\natexlab{}.
\newblock \showarticletitle{Chianti: a tool for change impact analysis of java
  programs}. In \bibinfo{booktitle}{\emph{Proceedings of the 19th annual ACM
  SIGPLAN conference on Object-oriented programming, systems, languages, and
  applications}}. \bibinfo{pages}{432--448}.
\newblock


\bibitem[\protect\citeauthoryear{Sawant, Robbes, and Bacchelli}{Sawant
  et~al\mbox{.}}{2018}]%
        {sawant2018reaction}
\bibfield{author}{\bibinfo{person}{Anand~Ashok Sawant}, \bibinfo{person}{Romain
  Robbes}, {and} \bibinfo{person}{Alberto Bacchelli}.}
  \bibinfo{year}{2018}\natexlab{}.
\newblock \showarticletitle{On the reaction to deprecation of clients of 4+ 1
  popular Java APIs and the JDK}.
\newblock \bibinfo{journal}{\emph{Empirical Software Engineering}}
  \bibinfo{volume}{23}, \bibinfo{number}{4} (\bibinfo{year}{2018}),
  \bibinfo{pages}{2158--2197}.
\newblock


\bibitem[\protect\citeauthoryear{Yu and Cheng}{Yu and Cheng}{2010}]%
        {YuCGRQS}
\bibfield{author}{\bibinfo{person}{Jeffrey~Xu Yu} {and}
  \bibinfo{person}{Jiefeng Cheng}.} \bibinfo{year}{2010}\natexlab{}.
\newblock \bibinfo{booktitle}{\emph{Graph Reachability Queries: A Survey}}.
\newblock \bibinfo{publisher}{Springer US}, \bibinfo{address}{Boston, MA},
  \bibinfo{pages}{181--215}.
\newblock


\bibitem[\protect\citeauthoryear{Zimmermann, Staicu, Tenny, and
  Pradel}{Zimmermann et~al\mbox{.}}{2019}]%
        {zimmermann2019small}
\bibfield{author}{\bibinfo{person}{Markus Zimmermann},
  \bibinfo{person}{Cristian-Alexandru Staicu}, \bibinfo{person}{Cam Tenny},
  {and} \bibinfo{person}{Michael Pradel}.} \bibinfo{year}{2019}\natexlab{}.
\newblock \showarticletitle{Small world with high risks: A study of security
  threats in the npm ecosystem}. In \bibinfo{booktitle}{\emph{28th
  $\{$USENIX$\}$ Security Symposium ($\{$USENIX$\}$ Security 19)}}.
  \bibinfo{pages}{995--1010}.
\newblock


\end{thebibliography}

\end{document}